\documentclass[lettersize,journal]{IEEEtran}
\IEEEoverridecommandlockouts
\usepackage{cite}
\usepackage{amsmath,amssymb,amsfonts}
\usepackage{algorithm, algorithmic}
\usepackage{graphicx, subcaption}
\usepackage{balance}
\usepackage{textcomp}
\usepackage{xcolor}
\usepackage{booktabs}

\def\BibTeX{{\rm B\kern-.05em{\sc i\kern-.025em b}\kern-.08em
    T\kern-.1667em\lower.7ex\hbox{E}\kern-.125emX}}
\begin{document}
\bstctlcite{IEEEexample:BSTcontrol}

\author{Sefa~Kayraklik,~\IEEEmembership{Graduate Student Member,~IEEE},
Ibrahim~Yildirim,~\IEEEmembership{Graduate Student Member,~IEEE},
        Ertugrul~Basar,~\IEEEmembership{Fellow,~IEEE},
        Ibrahim~Hokelek,~\IEEEmembership{Member,~IEEE}, and
        Ali Gorcin,~\IEEEmembership{Senior Member,~IEEE}
        \thanks{S. Kayraklik, I. Hokelek and A. Gorcin are with the Communications and Signal Processing Research (HİSAR) Lab., TÜBİTAK-BİLGEM, Kocaeli, Turkiye. (e-mail: \{sefa.kayraklik, ibrahim.hokelek, ali.gorcin\}@tubitak.gov.tr)}
        \thanks{S. Kayraklik, I. Yildirim, and E. Basar are also with the Communications Research and Innovation Laboratory (CoreLab), Department of Electrical and Electronics Engineering, Koç University, Sariyer, Istanbul, Turkey. (e-mail: ebasar@ku.edu.tr)}
        \thanks{I. Yildirim is also with the Faculty of Electrical and Electronics Engineering, Istanbul Technical University, Istanbul, Turkey, and also with the Department of Electrical and Computer Engineering, McGill University, Montreal, QC, Canada. (e-mail: yildirimib@itu.edu.tr)}
        \thanks{A. Gorcin is also with the Department of Electronics and Telecommunications Engineering, Istanbul Technical University, Istanbul, Turkey.}
}

\title{Practical Implementation of RIS-Aided Spectrum Sensing: A Deep Learning-Based Solution}

\markboth{IEEE Systems Journal}%
{Kayraklık \MakeLowercase{\textit{et al.}}: Practical Implementation of RIS-Aided Spectrum Sensing: A Deep Learning-Based Solution}

\maketitle

\begin{abstract}
This paper presents reconfigurable intelligent surface (RIS)-aided deep learning (DL)-based spectrum sensing for next-generation cognitive radios.
To that end, the secondary user (SU) monitors the primary transmitter (PT) signal, where the RIS plays a pivotal role in increasing the strength of the PT signal at the SU.
The spectrograms of the synthesized dataset, including the 4G LTE and 5G NR signals, are mapped to images utilized for training the state-of-the-art object detection approaches, namely Detectron2 and YOLOv7. By conducting extensive experiments using a real RIS prototype, we demonstrate that the RIS can consistently and significantly improve the performance of the DL detectors to identify the PT signal type along with its time and frequency utilization. This study also paves the way for optimizing spectrum utilization through RIS-assisted CR application in next-generation wireless communication systems.

\end{abstract}



\section{Introduction}
\IEEEPARstart{T}{he} existing electromagnetic radio spectrum is a valuable but limited natural resource that becomes insufficient due to the increasing number of wireless devices and applications. The current under-utilization of the licensed spectrum aggravates the problem and increases the demand for innovative and unconventional practical solutions \cite{Haykin_Sensing}. \textcolor{black}{While utilizing higher frequency bands with a large amount of available bandwidths is a longer-term solution, cognitive radio (CR), which can be directly applied to the licensed spectrum, is another prominent approach to address the spectrum under-utilization problem.} By exploiting spectrum sensing solutions, CR networks aim to provide reliable and fair communication for all users through dynamic spectrum access resulting in efficient and cost-effective spectrum utilization \cite{Haykin_CR}.

Reconfigurable intelligent surfaces (RISs) arise a key technology enabling dynamic control over the wireless propagation environment \cite{Basar_Access_2019, Wu_Tutorial}. A typical RIS consists of many passive reflecting elements that can be independently controlled to change the propagation characteristics of electromagnetic waves. By appropriately adjusting the phase shifts of the RIS elements, the incoming wireless signals can be conveyed towards the desired locations by facilitating a beneficial control over signal paths \cite{SimRIS_Mag}.

\begin{figure*}[t]
    \centering
    \includegraphics[width=0.95\linewidth]{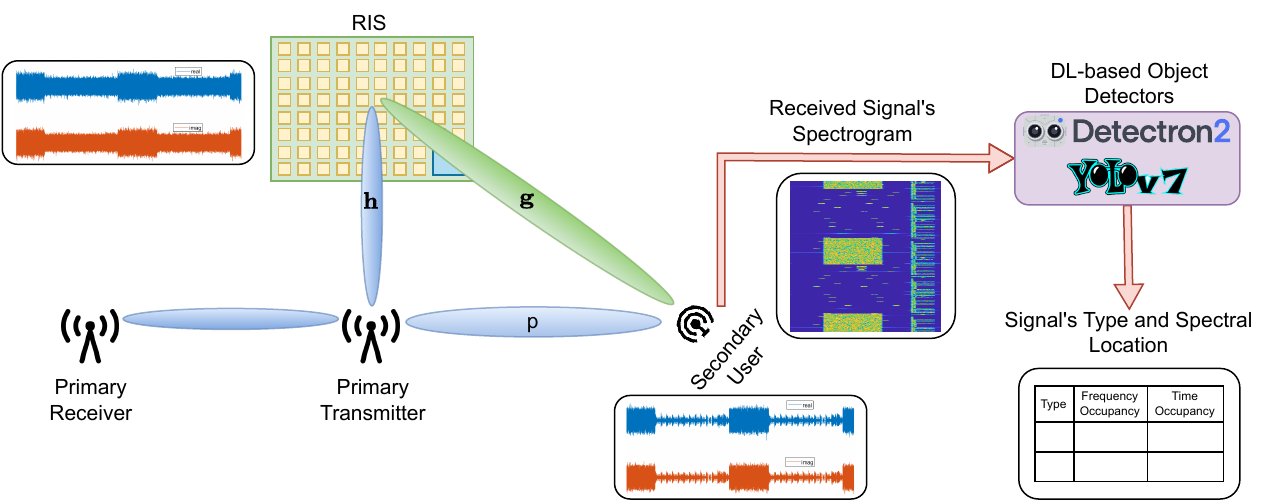} 
    \caption{{\color{black}The overview of the RIS-empowered spectrum sensing system with an DL-based solution.}} 
    \label{fig:SSSystemModel}
\end{figure*}

There has been growing interest in employing the RIS to enhance spectrum sensing capabilities and improve overall spectrum utilization by CR applications \cite{makarfi2021reconfigurable,wu2021irs,lin2022intelligent,nasser2022intelligent,ge2022ris, wu2023joint}.
In \cite{makarfi2021reconfigurable}, the authors demonstrate that the use of RIS configurations by the primary user (PU) can improve the spectrum sensing task of the secondary user (SU) in an RIS-assisted CR network. An RIS-enhanced energy detection framework for spectrum sensing is presented in \cite{wu2021irs}, where closed-form expressions for the average detection probability in the presence and absence of a direct link between the transmitter and receiver are derived. A weighted energy detection approach is proposed for the RIS-aided spectrum sensing system in the CR networks, where the RIS along with a passive beam-forming, is utilized to improve the received signal power of the SU \cite{lin2022intelligent}. In \cite{nasser2022intelligent}, the authors derive the detection probabilities when the RIS supports either the SU or the PU. Their results show that the RIS substantially improves the spectrum sensing performance for both cases. 
{\color{black} The maximum eigenvalue detection method is proposed for an RIS-empowered spectrum sensing system, in which the theoretical estimation for the number of the RIS reflecting elements is provided for accomplishing the desired detection probability \cite{ge2022ris}. In \cite{wu2023joint}, the authors proposed a new detection threshold regarding the false alarm rate to improve the performance of the spectrum sensing task in the RIS-assisted CR network where the RIS is optimized with two stages of sensing and transmission.}

{\color{black} The traditional signal processing methods such as adaptive thresholding, energy detection, matched filtering, and cyclostationary feature extraction for the spectrum sensing applications have inherent limitations, such as sub-optimal performance at low received signal powers, dependency on prior signal knowledge, and high computational complexity. In order to overcome these drawbacks, a set of studies considering deep learning (DL)-based spectrum sensing systems have been proposed \cite{prasad2020downscaled, kayraklik2022application, janu2022machine, syed2023deep}. In \cite{prasad2020downscaled}, the authors propose a framework with a downscaled faster RCNN to detect the signals and estimate their time-frequency positions in a wideband spectrum under interference of uninterested signals. The authors in \cite{kayraklik2022application} implement the DL-based object detectors for the wideband spectrum sensing problem by utilizing the signal's power spectral densities. \cite{janu2022machine} and \cite{syed2023deep} present comprehensive literature reviews for the spectrum sensing application with machine learning and deep learning approaches, respectively.
To the best of our knowledge, these DL-based studies for spectrum sensing applications do not utilize the RIS in the system model. 
}

{\color{black}
In this paper, we propose the first-ever practical implementation of the RIS-empowered DL application for spectrum sensing employing a real RIS prototype. The communication between the primary transmitter (PT) and the SU occurs through the combination of a direct link and the reflected signals from the RIS. Assuming that the horn antennas of the PT and the SU remain non-intersecting, resulting in the reception of the reflected signal from the PT only through the RIS-assisted path. Therefore, an RIS is utilized to enhance the spectrum sensing capacity of the SU by improving the SU's received signal power from the PT. The spectrogram of the received signal at the SU is calculated by taking the squared magnitude of the short-time Fourier transform (STFT) of the signal. Then, each spectrogram is converted to a red-green-blue (RGB) image with a specific colormap corresponding to the power values of the spectrogram points. In our proposed spectrum sensing method,} the state-of-the-art object detection models, namely Detectron2 \cite{wu2019detectron2} and "You Only Look Once" (YOLO)v7 \cite{wang2023yolov7}, are trained using the spectrogram images of the synthesized dataset including the fourth-generation long-term evolution (4G LTE) and fifth-generation new radio (5G NR) signals. Utilizing the spectrogram images, the spectrum sensing task can be treated as object detection for estimating the type of the signal and its spectral usage. \textcolor{black}{ The measurement experiments employing the RIS prototype and software-defined radios (SDRs) demonstrate remarkable enhancements in the performance of the DL-based spectrum sensing method by accurately characterizing the PT signal type and its temporal and spectral information. These results indicate that the RIS technology can be effectively employed in the CR system in such a way that the SU can utilize the unoccupied spectrum of the PT by precisely detecting the PT's spectrum activities.} 

{\color{black} The organization of the paper is as follows. Section II describes the system and signal models. The DL-based spectrum sensing approach is presented in Section III, along with the state-of-the-art object detectors, the training parameters, and dataset generation. Section IV provides the detectors' performance analysis through measurement experiments. Finally, Section V concludes our work on the RIS-aided spectrum sensing system.}

\textcolor{black}{\textit{Notation:} The following notations are used throughout this paper. Bold upper/lower case letters denote matrices/vectors. $\mathbb{C}^{N}$ represents the space of $N \times 1$ complex vectors. $(.)^H$ denotes the conjugate transpose of a vector or matrix. diag\{$\mathbf{x}$\} represents a diagonal matrix with the diagonal elements given by entry of vector $\mathbf{x}$. The notation $x \sim \mathcal{N}_{\mathbb{C}}\left(\mu,\sigma^2\right)$ is used to represent that $x$ is a complex Gaussian random variable with mean $\mu$ and variance $\sigma^2$. We also use the floor function, denoted as $\lfloor x\rfloor$, which maps a real number $x$ to the greatest integer less than or equal to $x$.}

\section{System Model}
Fig. \ref{fig:SSSystemModel} shows an RIS-empowered spectrum sensing system, where an RIS is employed to assist the SU in acquiring the parameters of the signal broadcasted by the PT. The communication between the PT and the PR takes place through a line-of-sight (LoS) link, assuming that the reflected signal from the RIS to the PR can be neglected compared to the LoS signal {\color{black} since the RIS is positioned near the SU}. The communication between the PT and SU occurs through a direct link and the reflected signals from the RIS. The main objective of the RIS is to maximize the received signal power by the SU through the RIS. 

Here, we incorporate an RIS consisting of $N=N_x\times N_y$ reflecting elements arranged in a uniform planar array, visually represented by yellow squares in Fig. \ref{fig:SSSystemModel}. 
$\mathbf{h}\in \mathbb{C}^{N}$, $\mathbf{g}\in \mathbb{C}^{N}$, and $p\in \mathbb{C}$ represent the channels of the PT-RIS, RIS-SU, and PT-SU, respectively.
Here, $x[k]$ and $r[k]$ are the complex baseband signals transmitted by the PT and received by the SU, respectively, while $n[k]\sim \mathcal{N}_{\mathbb{C}}\left(0,\sigma_{n}^2\right)$ denotes the noise at the SU. 
{\color{black} The SU receives either the reflected form of $x[k]$ from the RIS when the PT is in transmit mode or experiences noise when the PT is in idle mode.}
The received signal, $r[k]$, at the SU under the hypotheses $\mathcal{H}_0$ (the PT is in idle mode) and $\mathcal{H}_1$ (the PT is in transmit mode) can be expressed as follows
\begin{equation}
\begin{aligned} \label{eq:received_ss}
   \mathcal{H}_0: r[k] &= n[k], \\
   \mathcal{H}_1: r[k] &= \left(\mathbf{g}^H\mathbf{\Theta} \mathbf{h} + p\right)x[k]+n[k], 
\end{aligned}
\end{equation}
where $\mathbf{\Theta}=\text{diag}\left\{\alpha_{1} e^{j \phi_{1}}, \cdots, \alpha_{N} e^{j \phi_{N}}\right\}$ represents the phase shift configuration matrix of the RIS. $\alpha_{n}$ and $\phi_{n}$ denote the magnitude response and phase shift of the $n$th element, respectively. 

{\color{black} The DL-based object detectors utilize the spectrogram images of the received signal, whose flow is depicted in Fig. \ref{fig:SSSystemModel}, in order to estimate the PT's signal type and its temporal and spectral resource occupancy. The spectrogram of the received signal, $r[k]$, is computed by taking the squared magnitude of the STFT of the signal \cite{boashash2015time}. The STFT analyzes the Fourier transform of a short-period portion of the whole signal by utilizing a window function. The STFT of the received signal is calculated by moving the window function, $v[k]$, of length $S_v$ over the received signal, $r[k]$, of length $S_r$ with the hop intervals of $M$ samples and computing the discrete Fourier transform (DFT) for each portion of the windowed signal. The $m$th row of the STFT matrix, $R_m[\omega]$, containing the windowed signal's DFT centered around time $m\times M$ is calculated as follows
\begin{equation}
\begin{aligned}
    & R_m[\omega] = \sum_{k=0}^{S_v-1} r[k] v[k-mM] e^{-j\frac{2\pi}{S_v} \omega k}, \\
    & \mathbf{R} = 
    \begin{bmatrix}
        R_1[0] & R_1[1] & \dots & R_1[S_v-1]\\
        R_2[0] & R_2[1] & \dots & R_2[S_v-1]\\
        \vdots & \vdots & \ddots & \vdots\\
        R_c[0] & R_c[1] & \dots & R_c[S_v-1]
    \end{bmatrix},
\end{aligned}
\end{equation}
where $c = \left \lfloor \frac{S_r - S_v}{M} \right \rfloor$ is the total number of rows and $\mathbf{R}$ is the whole STFT matrix. Therefore, the spectrogram matrix, $\mathbf{S}$, of the received signal is derived as the following expression
\begin{equation}
\mathbf{S}_{i,j} = |\mathbf{R}_{i,j}|^2, \quad \forall i \in [1,c]\text{ and } \forall j \in [1, S_v],  
\end{equation}
where each element of the spectrogram is equal to the squared magnitude of the corresponding STFT element. The subscripts of the matrices, $i$ and $j$, denote row and column numbers, respectively. Since the proposed spectrum sensing approach utilizes the received signal's spectrograms as images, the calculated spectrograms are mapped into a RGB image with a specific colormap corresponding to the power values of the spectrogram points. The dB-scaled values of the spectrogram matrices are converted to RGB images by mapping the quantized values of the spectrograms between $[0,1]$ to a specific colormap with 8 bits. The conversion of the dB-scaled spectrogram matrices to RGB image matrices is given as follows
\begin{equation}
\begin{aligned}
   \mathbf{RGB} &= map\left \{ \frac{\mathbf{S}_{dB}-\min(\mathbf{S}_{dB})}{\max(\mathbf{S}_{dB})-\min(\mathbf{S}_{dB})} \right \} ,
\end{aligned}
\end{equation}
where $\mathbf{S}_{dB}$ is equal to $10\log_{10}\mathbf{S}$ and the transform of $map\{ \dots \}$ converts the indexed values between $[0,1]$ to specific colors according to their magnitudes by quantizing them with 8 bits associated with a given colormap. For example, MATLAB's parula colormap represents the values $[0,1]$ starting from blue to yellow with $256$ different colors.
 
When the behavior of the received signal is analyzed, the corresponding spectrogram under the hypothesis of $\mathcal{H}_0$ is more likely to have one color across the time and frequency axes. On the other hand, the spectrogram generated under the hypothesis of $\mathcal{H}_1$ contains brighter colors for the occupied spectrum regions due to the high signal power, whereas the generated spectrogram has darker colors for unoccupied areas since they are on the noise floor. Therefore, by treating the PT's signals and the unoccupied bands as objects on the spectrogram images, the state-of-the-art DL-based object detectors of Detectron2 and YOLOv7 can be utilized to perform the spectrum sensing task in the SU.
}

\section{DL-based Spectrum Sensing}
CR networks aim to maximize spectrum efficiency by dynamically detecting the PT's transmission activity, including the signal type and spectral characteristics through the spectrum sensing application and utilizing the unoccupied spectrum resources for the SU. Traditional signal processing methods for spectrum sensing application includes adaptive thresholding, energy detection, matched filtering, and cyclostationary feature extraction. These methods experience inherent limitations, such as sub-optimal performance at low received signal powers, dependency on prior signal knowledge, and high computational complexity. Nevertheless, next-generation wireless networks accommodate massive machine-type communication, where narrow-band network elements will have limited signal processing and transmit power capabilities. DL-based solutions can be utilized to overcome the deficiencies of traditional spectrum sensing approaches {\color{black} by rapid adaptation of learned models to different spectrum conditions. In fact, DL-based approaches offer more flexible detection and estimation of the signal types than traditional methods, especially in dynamic wireless environments and changing spectrum usage scenarios. Note that DL-based methods require large amounts of labeled data and high computational power in the model training phase for higher performance. It might have limited interpretability of the model’s decision-making process.}

A set of known signal types can be detected and identified by utilizing the spectrogram of the received signals as images, where the state-of-the-art object detection approaches can be directly applied. As the detection performance of a spectrum sensing application ultimately depends upon the strength of the received signal, the RIS can play a pivotal role in increasing the PT signal power received by the SU. {\color{black} Therefore, incorporating the RIS into the spectrum sensing system is anticipated to provide notable improvements in the spectrum sensing performance.}

\begin{figure}[t]
    \centering
    \includegraphics[width=0.99\linewidth]{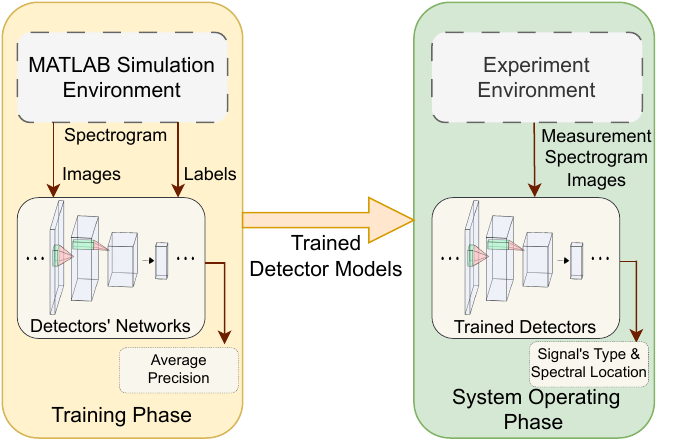}
    \caption{{\color{black}The workflow of the phases of training and system operation of detectors for DL-based spectrum sensing.}}
    \label{fig:SSWorkflow} 
\end{figure}

{\color{black} Fig. \ref{fig:SSWorkflow} illustrates the workflow of the proposed DL-based spectrum sensing consisting of training and system operation phases. In the training phase, the dataset generated using the MATLAB simulation environment is fed to the object detector models to be trained. The simulated dataset includes the spectrogram images and signal labels corresponding to the signal's type and spectral locations of unoccupied bands, 4G LTE, and 5G NR. During the training phase, the detector models' training parameters are determined according to one of the performance metrics of average precision of the estimated signal's types and spectral locations. After the completion of the training phase, the successfully trained detector models are transferred to the system operation phase. The measurement spectrogram images from the experiment environment are assigned trained detectors of Detectron2 and YOLOv7 to infer the images during the system operation phase. The detector models report the estimated types and spectral locations of the detected signals to the SU in order to inspect the PT's activities. If the spectrum is not occupied by the PT, the SU will utilize the empty spectrum, resulting in more effective spectrum usage.
} 

\subsection{State-of-the-Art Object Detectors: Detectron2 and YOLOv7}
{\color{black} Object detection can be considered a prominent domain in computer vision and pertains to the task of accurately locating and classifying objects in a given target image. This field finds extensive application in diverse areas, from classifying and locating enemy threads in synthetic aperture radar images to detecting cancer cells in magnetic resonance imaging scans. 
This subsection aims to provide a methodology for the detection, classification, and localization of the PT signals on images that illustrate the frequency and time domain recordings of the signals. }
In this study, we utilize two prominent DL-based object detector approaches: Detectron2 \cite{wu2019detectron2} and YOLOv7 \cite{wang2023yolov7}. 

Detectron2, which is developed by Facebook AI Research (FAIR) as an open-source library, provides a modular framework for rapid implementation and assessment of forthcoming research in the computer vision area. The object detection and segmentation methods of Detectron2 include Faster R-CNN, Mask R-CNN, TensorMask, PointRend, RetinaNet, and DensePose. 

YOLOv7, {\color{black} which is a revised version of YOLO \cite{redmon2016yolo},} allows the target image to be handled as a network input and directly returns the bounding box probabilities and locations of the detected objects. {\color{black} YOLO is a neural network consisting of a single end-to-end structure, which divides each target image into grids to detect objects when their centers intersect a grid cell.} It is reported in \cite{wang2023yolov7} that YOLOv7 outperforms in terms of the inference speed and accuracy thanks to its re-parameterization of the model network and dynamic label assignment. 

\begin{figure*}[t]
\centering
\subfloat[]{\label{fig:SSspectrum:a} 
\includegraphics[width=0.23\linewidth]{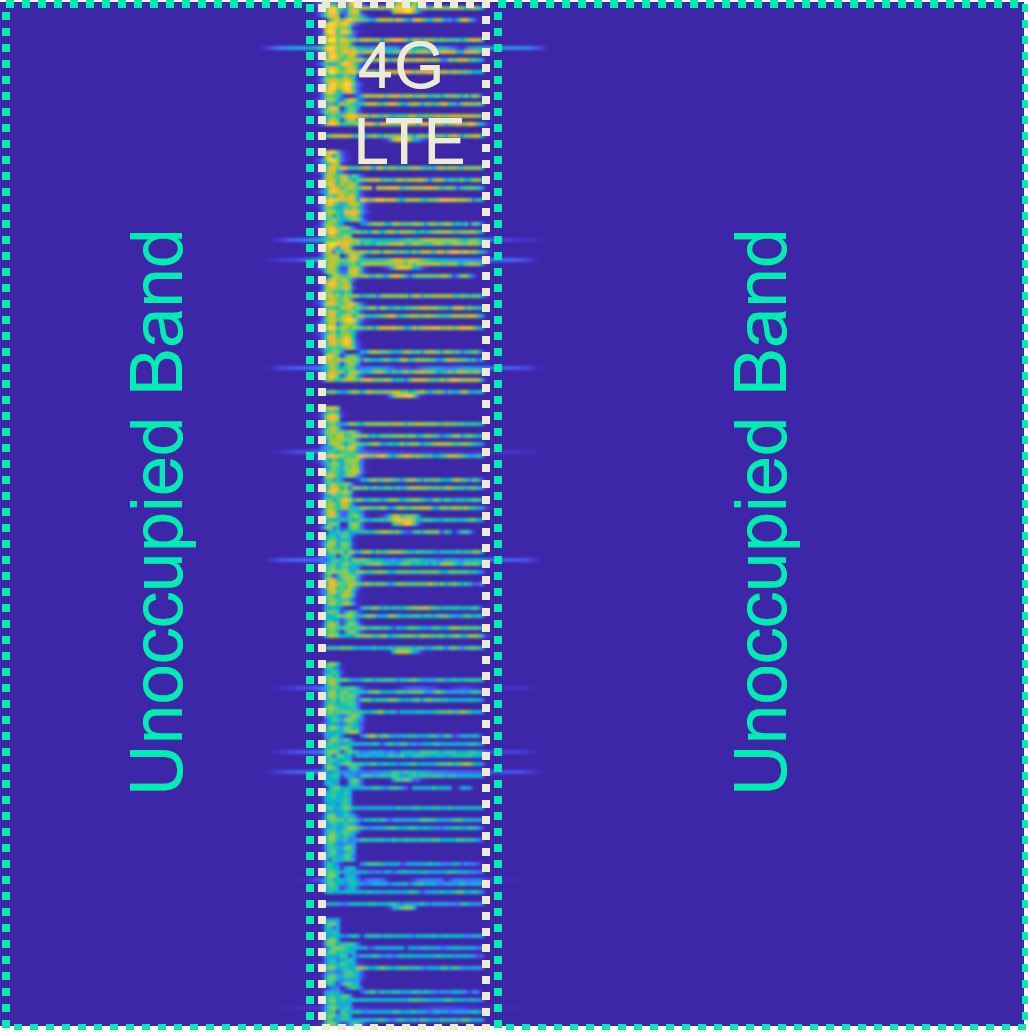}}
\subfloat[]{\label{fig:SSspectrum:b} 
\includegraphics[width=0.23\linewidth]{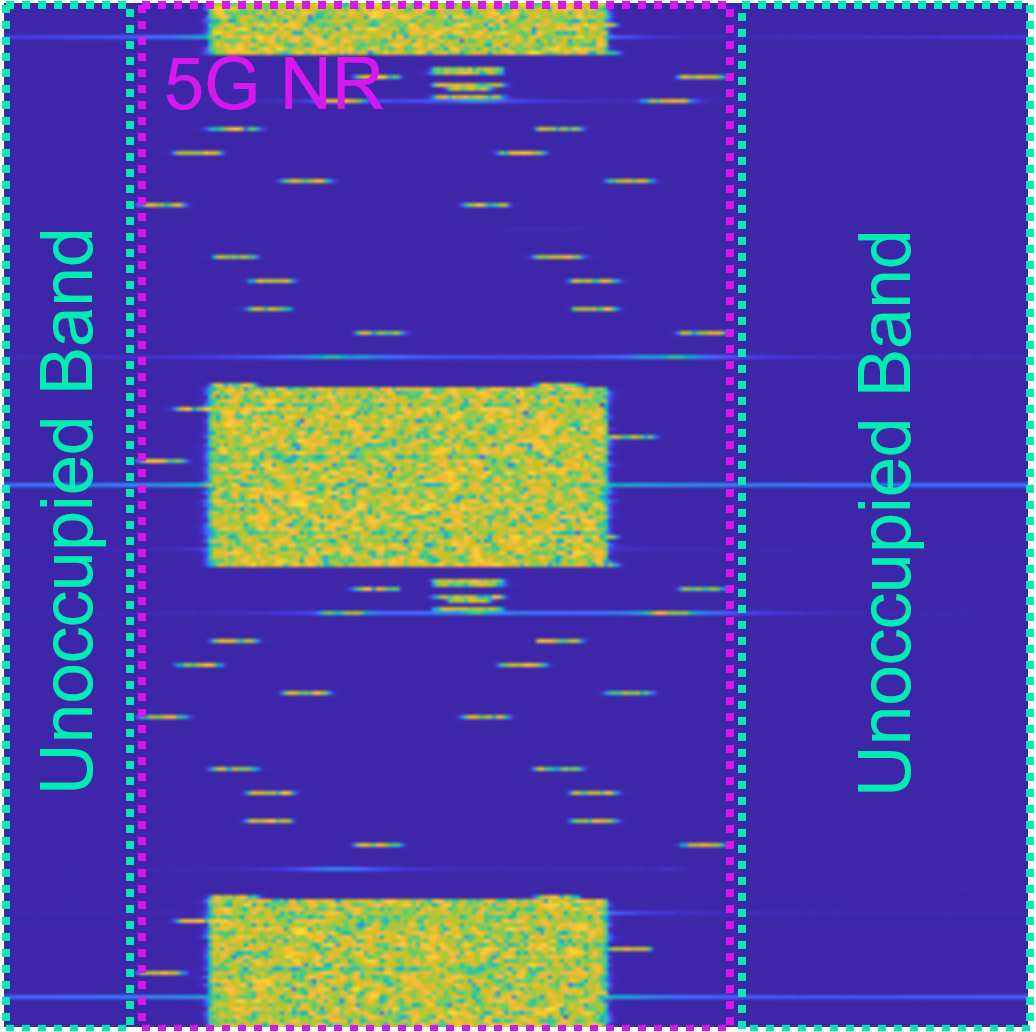}}
\subfloat[]{\label{fig:SSspectrum:c} 
\includegraphics[width=0.23\linewidth]{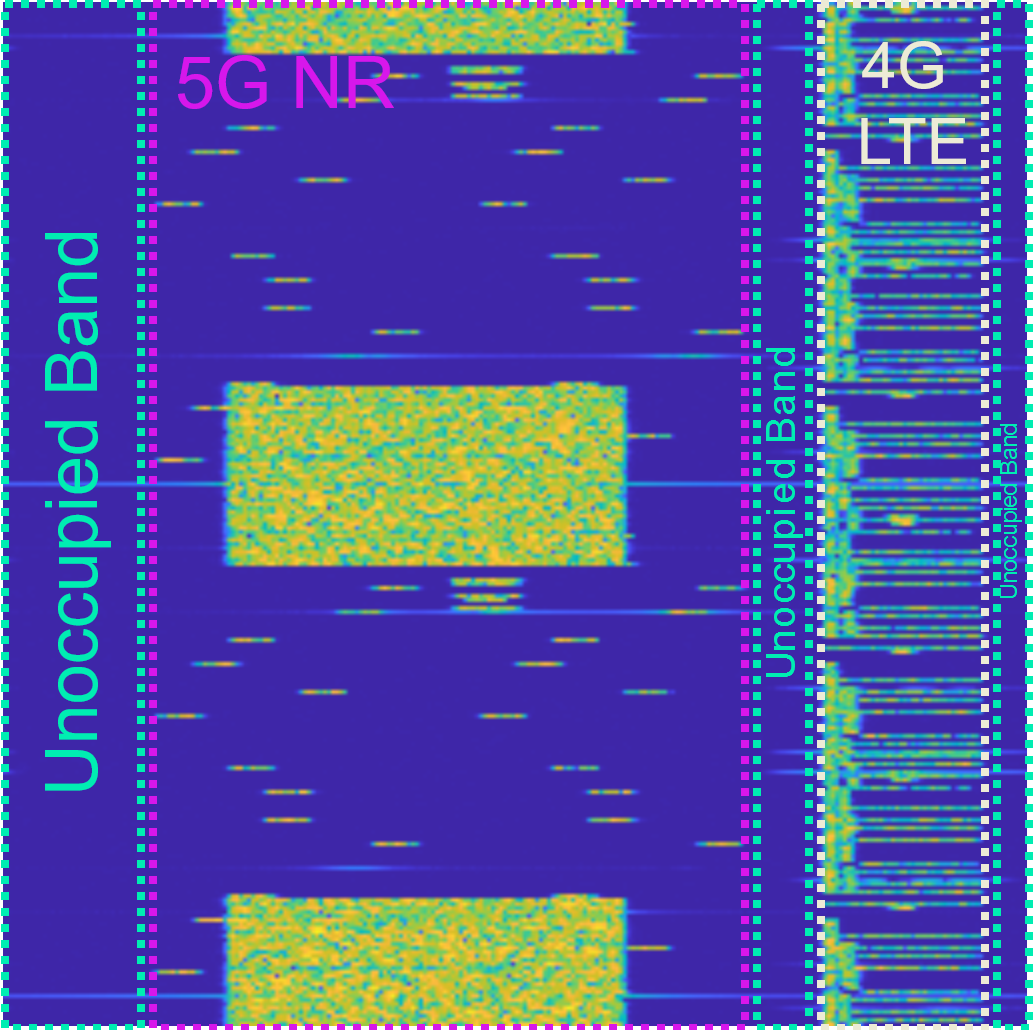}}
\caption{The spectrograms of the generated signals of (a) 4G LTE, (b) 5G NR, and (c) both 4G LTE and 5G NR, where the x-axis and y-axis represent the frequency and time domains, respectively.}
\label{fig:SSspectrum}
\end{figure*}

\subsection{Training Parameters of the Detectors}
The training parameters of YOLOv7\footnote{\label{FN_yolov7}Available at: https://github.com/WongKinYiu/yolov7} and Detectron2\footnote{\label{FN_detectron}Available at: https://github.com/facebookresearch/detectron2} models, whose codes are publicly available, are experimentally determined for the spectrum sensing application. {\color{black} These highly complex DL-based detector models require a large dataset for achieving sufficient model accuracy since there is a large number of parameters to be
trained. Therefore, } the transfer learning approach is employed to reduce the training effort for the spectrum sensing models of both detectors, in which the pre-trained weights are utilized as the initial weights. The Common Objects in Context (COCO) dataset is utilized to obtain these pre-trained weights. This transfer learning approach significantly reduces the number of epochs required for training the models. {\color{black} Furthermore, this method is particularly helpful in the field of spectrum sensing, where labeled data is rare and costly to produce due to its ability to convey general features of the models pre-trained on large data sets to a new task. In fact, this significantly reduces the amount of data required and accelerates the training process. The transfer learning approach also helps to avoid overfitting, especially when working with small datasets, and improves the generalization of models to new, unseen data, which is critical for the robust detection of various types of signals in CR systems. }

The pre-trained model of YOLOv7\textsuperscript{\ref{FN_yolov7}} in \cite{wang2023yolov7} is utilized with {\color{black}the hyperparameters given in the file of \textit{hyp.scratch.p6.yaml}\textsuperscript{\ref{FN_yolov7}}}, the batch and epoch sizes of $64$ and $200$, respectively. For Detectron2, the mask R-CNN architecture with the pre-trained model of \textit{mask\_rcnn\_X\_101\_32x8d\_FPN\_3x}\textsuperscript{\ref{FN_detectron}} in \cite{wu2019detectron2} is used. Additionally, Detectron2 is trained with {\color{black}the hyperparameters given in the file of \textit{Base-RCNN-FPN.yaml}\textsuperscript{\ref{FN_detectron}}} a learning rate of $0.0025$, an image per batch of $16$, a batch size per image of $512$, and $4000$ iterations with every $20$ iteration corresponding to an epoch. {\color{black} The training and testing of the detectors are performed using a server with an Intel Xeon Platinum processor, an Nvidia Tesla V100S graphics processing unit, and a random access memory of 64GB.}

\subsection{Signals of Interest, Dataset Representation and Generation}\label{subsec:signals}
We consider the following three scenarios to train the detectors used for the RIS-assisted DL-based spectrum sensing application: The PT transmits \textit{(i)} 4G LTE, \textit{(ii)} 5G NR, and \textit{(iii)} both 4G LTE and 5G NR signals simultaneously. These scenarios are generated using MATLAB's LTE and 5G NR Toolboxes, which are compliant with the 3GPP standards. The spectrograms of the generated signals, calculated by taking a sampling rate of $60$ MHz at the center frequency of $5.2$ GHz and a signal duration of $40$ ms, are illustrated in Fig. \ref{fig:SSspectrum}, where each spectrogram is saved as a JPG image having $256\times 256$ pixels. Each spectrogram is processed using a Hanning window with an overlapping ratio of $10\%$ and an FFT size of $4096$. The rows and columns in Fig. \ref{fig:SSspectrum} indicate the frequency and time domains of the signals, respectively. Additionally, the colormap represents the signal strength, with blue representing a low signal power and yellow representing a high signal power. {\color{black} Note that in Fig. \ref{fig:SSspectrum}, example spectrogram images of the signals of interest are provided with a high signal power in order for the signals to be observed more clearly.}

\begin{figure*}[t]
\centering
\subfloat[]{\label{fig:SSSimResults:a} 
\includegraphics[width=0.27\linewidth]{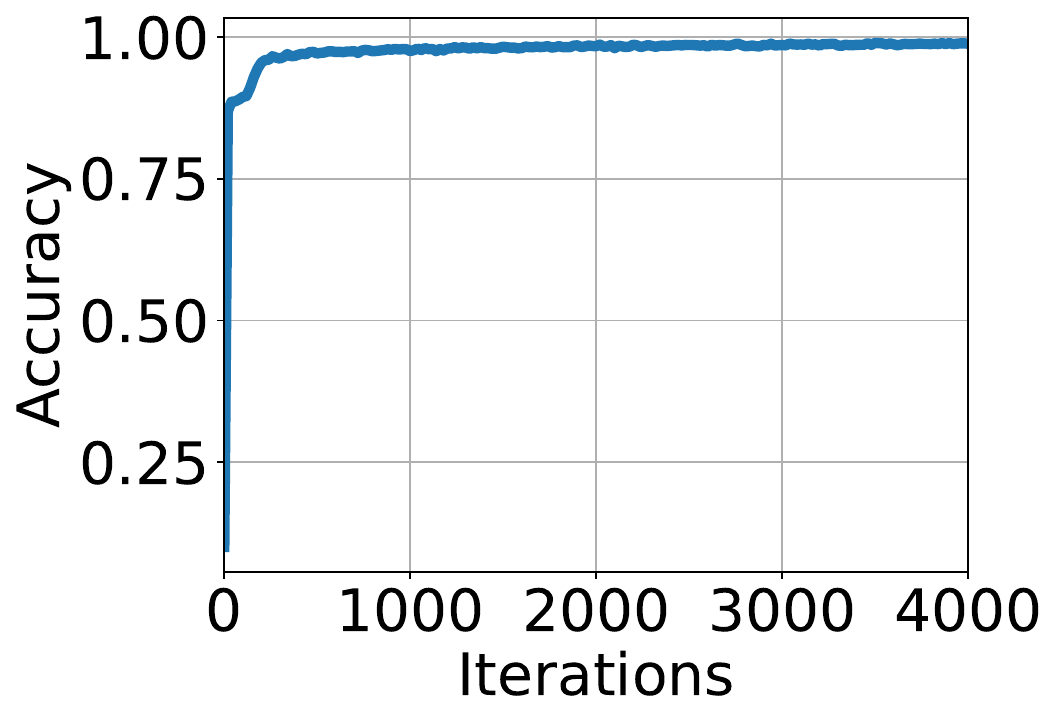}}
\subfloat[]{\label{fig:SSSimResults:b} 
\includegraphics[width=0.25\linewidth]{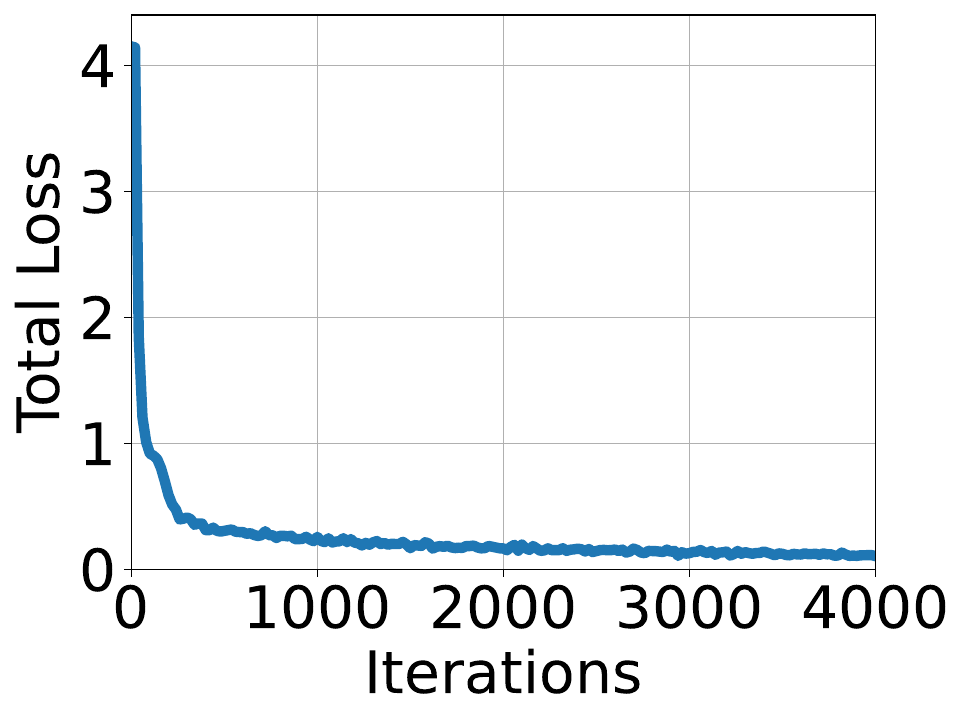}}
\subfloat[]{\label{fig:SSSimResults:c} 
\includegraphics[width=0.27\linewidth]{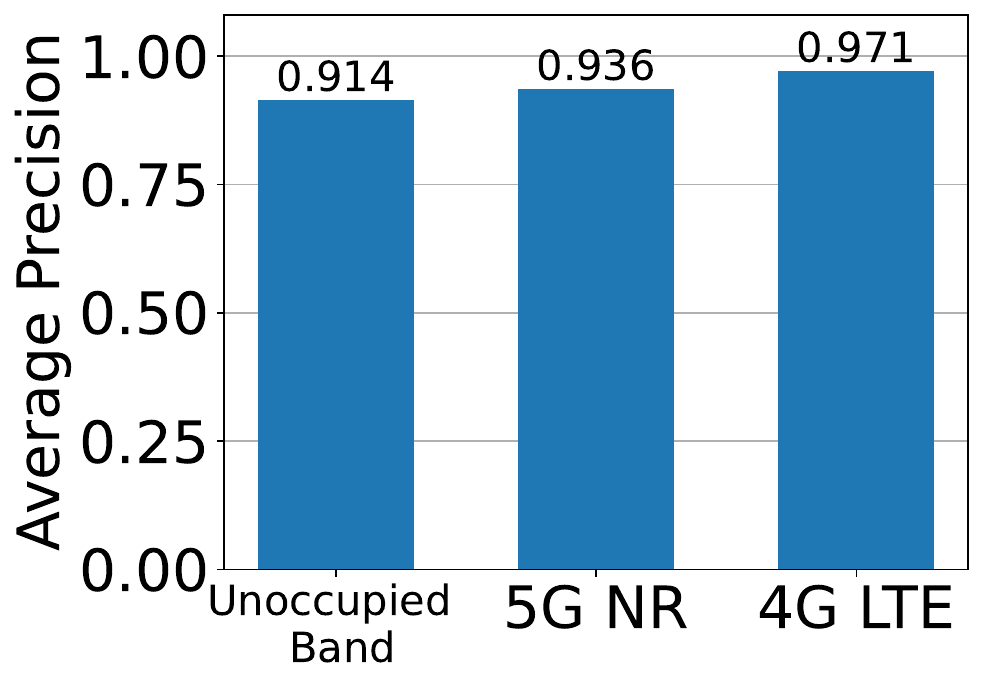}}\\[10pt]
\subfloat[]{\label{fig:SSSimResults:d} 
\includegraphics[width=0.27\linewidth]{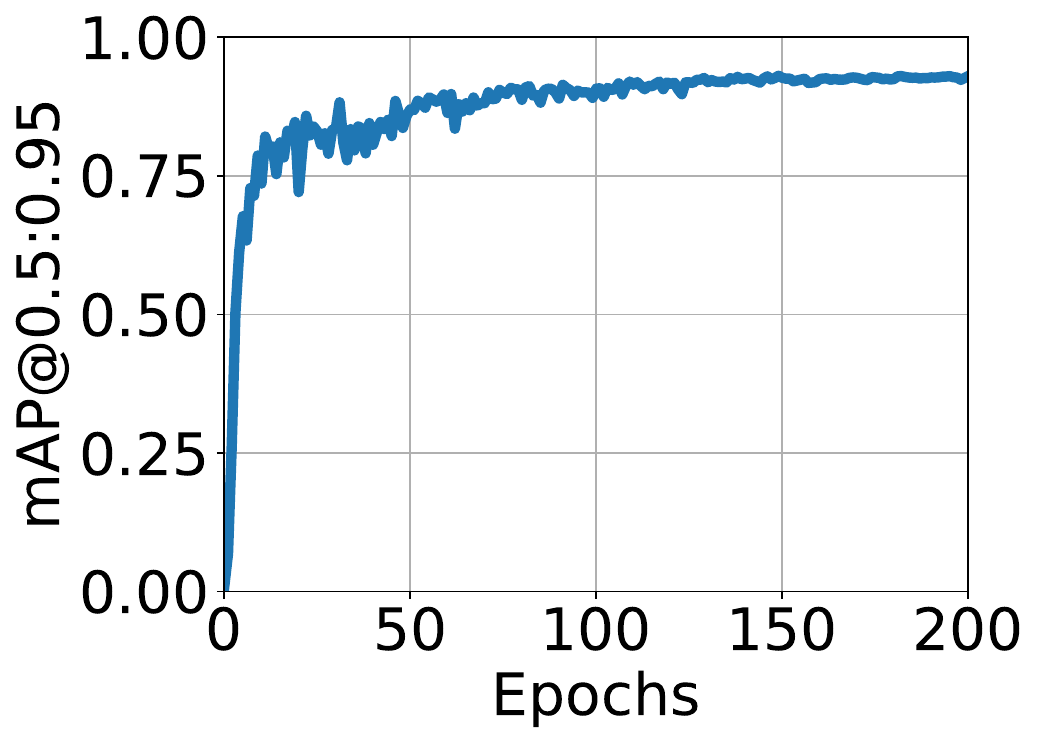}}
\subfloat[]{\label{fig:SSSimResults:e} 
\includegraphics[width=0.265\linewidth]{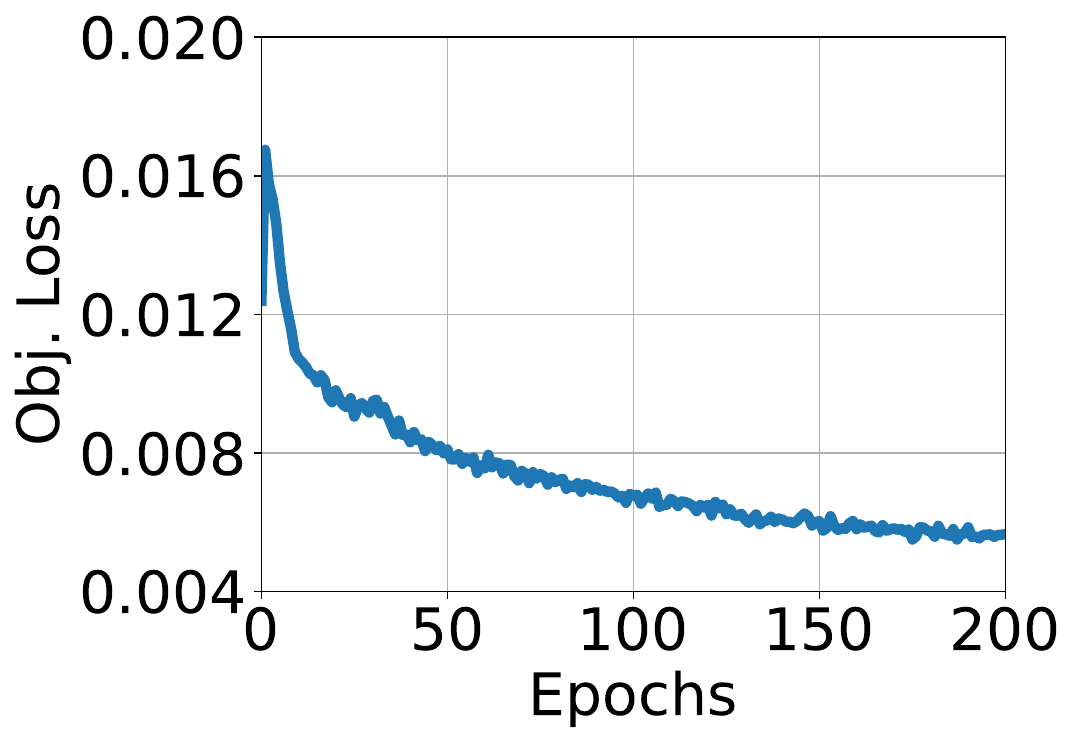}}
\subfloat[]{\label{fig:SSSimResults:f} 
\includegraphics[width=0.27\linewidth]{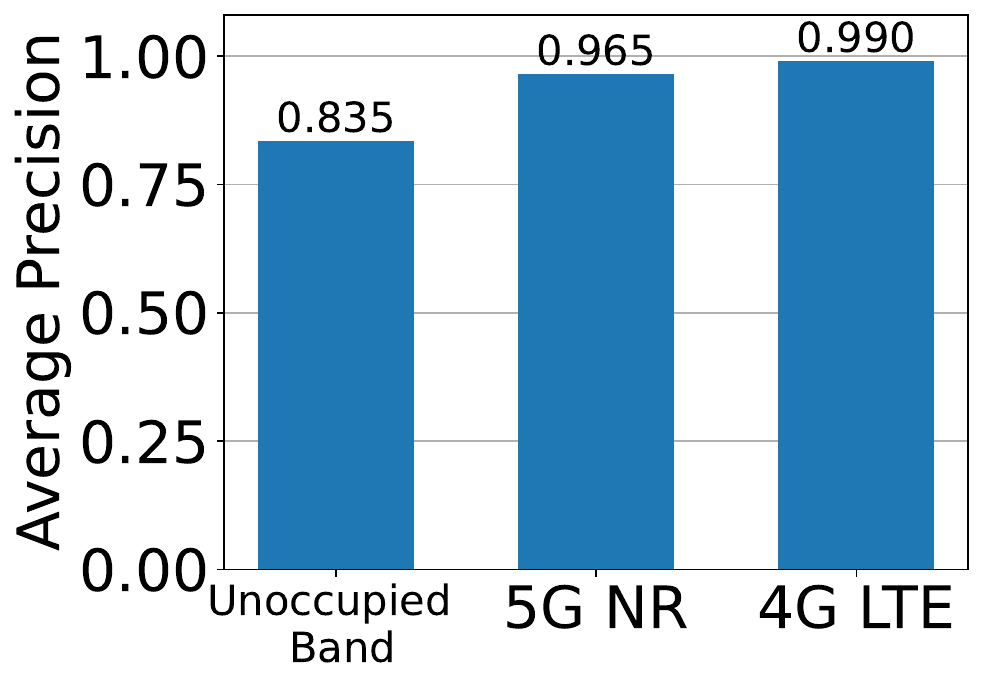}}
\caption{Detectron2's (a) accuracy, (b) total loss, and (c) average precision with the test dataset; YOLOv7's (d) mAP@0.5:0.95, (e) objectness loss, and (f) average precision with the test dataset.}
\label{fig:SSSimResults} 
\end{figure*}

{\color{black} A synthesized dataset consisting of signals of the PT's transmission is generated through MATLAB's toolboxes in order to train the object detectors, assembled by the architectures of Detectron2 and YOLOv7.} 
The parameters for the 4G LTE signals are chosen as frequency division duplexing and the bandwidth of $\{ 5,10,15,20\}$ MHz, whereas the parameters for the 5G NR signals are selected as the subcarrier spacing of $\{15,30\}$ kHz and the bandwidth of $\{ 10,15,20,25,30,40,50\}$ MHz. In addition, the target signal-to-noise ratio (SNR) values of $\{0,20,50\}$ dB and the target Doppler values of $\{0,10,500\}$ Hz are set for the channel conditions. For each signal transmission scenario, $900$ different signals for training and $300$ signals for testing are generated with MATLAB's simulated channels. Therefore, there are  $2700$ spectrogram images to be fed to the detectors to train and $900$ spectrogram images to test the trained detectors for all three scenarios. Note that these signals are generated using the example code in Mathworks website \cite{mathworks}. Instead of the SNR values of $\{-10,0,50\}$ dB, we utilize $\{0,20,50\}$ dB since the RIS increases the received signal power at the SU by focusing the reflected signals towards the SU. This ensures that the detectors are trained using a more suitable dataset. 

The training dataset of $2700$ spectrograms is fed to Detectron2 and YOLOv7 models to train their weights. These detectors are utilized to identify the 4G LTE, 5G NR, and unoccupied bands on spectrogram images. Figs. \ref{fig:SSSimResults}(a) and \ref{fig:SSSimResults}(b) illustrate the accuracy of Detectron2 and its total loss with respect to the number of training iterations, respectively. As the number of iterations increases, the accuracy rapidly approaches towards unity while the total loss decreases towards zero. The total loss is the weighted sum of the classification, localization, and mask losses. The average precision (AP) values of the trained Detectron2 model, which are obtained by the test dataset including $900$ spectrogram images, are shown in Fig. \ref{fig:SSSimResults}(c). The AP values of the unoccupied band, 5G NR, and 4G LTE are higher than $0.91$, and the highest AP of $0.971$ is achieved for 4G LTE. The SU can transmit its signals when there are no 5G NR and 4G LTE signals of the PT. In addition, the SU can transmit its signals using only unoccupied regions of the spectrum when there are 5G NR and 4G LTE signals.

For the YOLOv7 detector's training process, the mean AP (mAP), which is calculated at the intersection over union (IoU) from $0.50$ to $0.95$ with the step size of $0.05$ (mAP@0.5:0.95), is shown in Fig. \ref{fig:SSSimResults}(d). Note that the IoU measures the overlapping of predicted signal boundaries with the real signal boundaries. The objectness loss through the epochs is shown in Fig. \ref{fig:SSSimResults}(e). As the number of training epochs increases, the model precision improves while the objectness loss decreases. The AP values of the trained YOLOv7 model, which are obtained by the test dataset including $900$ spectrogram images, are shown in Fig. \ref{fig:SSSimResults}(f). The results show that YOLOv7 can detect 4G LTE and 5G NR signals more accurately, while Detectron2 can identify unoccupied bands better. Note that these trained models will be utilized in practical measurements, including the RIS, in the following section. The test results indicate that the models are successfully trained. However, it is suggested to use the results of practical measurements to compare the performance of the models.

\section{Performance Analysis of the Detectors in Practical Measurement}
\begin{figure}[t]
    \centering
    \includegraphics[width=0.95\linewidth]{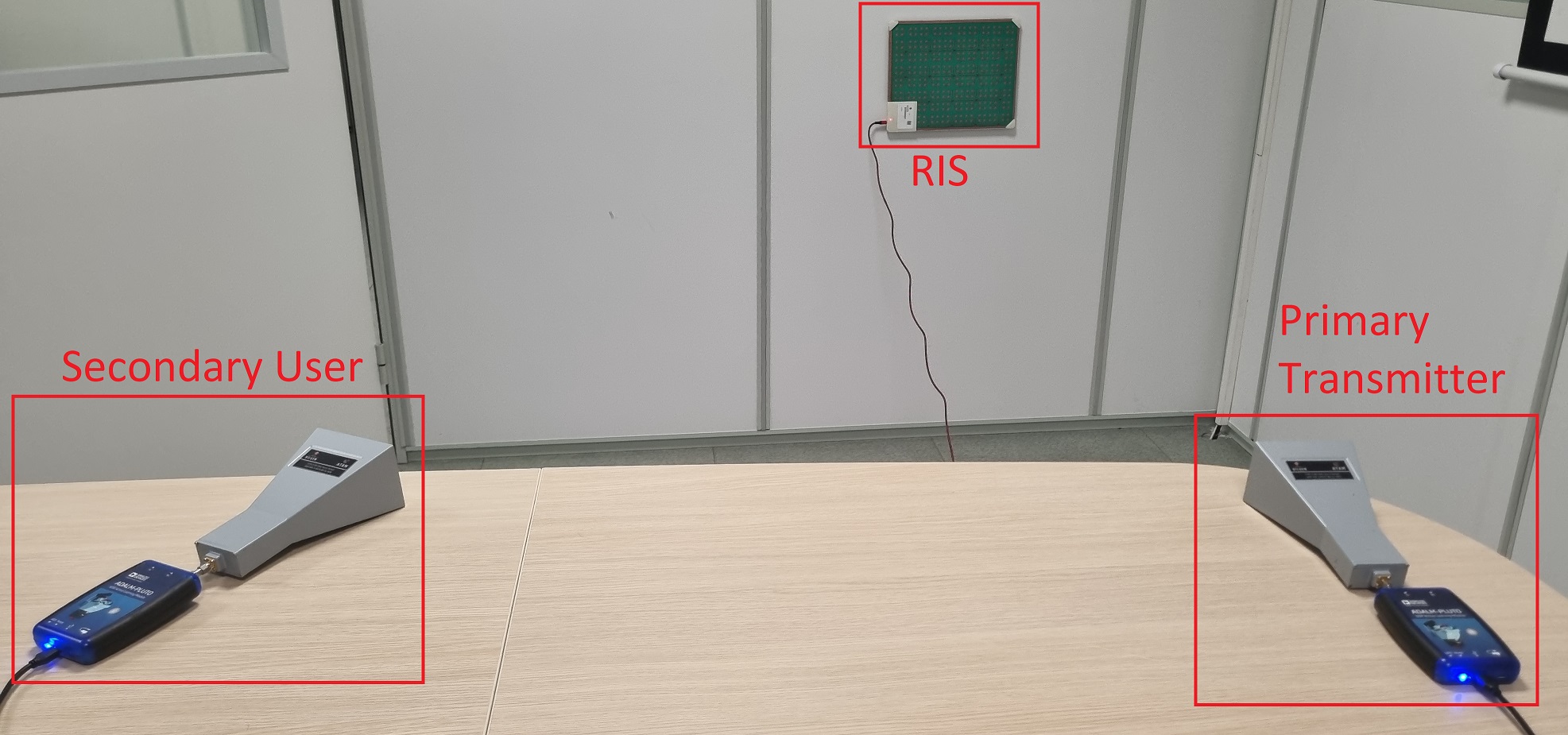}
    \caption{The RIS-aided spectrum sensing measurement setup.}
    \label{fig:SSSchematic} 
\end{figure}

The measurement setup of the RIS-enhanced spectrum sensing system is illustrated in Fig. \ref{fig:SSSchematic}. The experiments are conducted with ADALM-PLUTO SDR modules to realize the over-the-air transmission and reception by the PT and SU, respectively. The PT and SU are located to ensure that their horn antenna beams remain non-intersecting, resulting in the reception of the reflected signal from the PT only through the RIS-assisted path. The Greenerwave RIS prototype \cite{greenerwave} operating at $5.2$ GHz is located on the wall to boost the power of the received signals by the SU by adjusting the phase shifts of the reflecting elements. The RIS is formed in the structure of a uniform planar array with $8 \times 10$ reflecting elements, but its left bottom corner of $2\times 2$ region is designated for its controller, ending up with a total of $76$ reflecting elements. Each element's phase shift can be individually adjusted by two PIN diodes, which correspond to the horizontal and vertical polarization, as $0^\circ$ and $180^\circ$, resulting in four different states. {\color{black} The summary of hardware utilized in the measurement experiments of the RIS-assisted spectrum sensing system is given in Table \ref{table:specs}}. 

\begin{table}[t]
\centering
\caption{{\color{black}The hardware specifications of the experiment setup.}}
\label{table:specs}
\resizebox{0.48\textwidth}{!}{%
\begin{tabular}{@{}ll@{}}
\toprule
\textbf{Hardware} &  \multicolumn{1}{c}{\textbf{Description}} \\ \midrule
Transceiver         &    \begin{tabular}[c]{@{}l@{}}ADALM-PLUTO SDR, which is modified to operate at \\ $5.2$ GHz with an instantaneous bandwidth of $60$ MHz.\end{tabular}       \\[8pt]
RIS         &        \begin{tabular}[c]{@{}l@{}}Greenerwave prototype operating at $5.2$ GHz with the \\ $76$ reflecting elements whose phase shifts are controlled \\ as $0^\circ$ and $180^\circ$ for both polarization.\end{tabular}      \\[13pt]
Antenna         &    \begin{tabular}[c]{@{}l@{}}Horn antennas with $13$ dBi directivity around the frequ-  \\ ency of $5$ GHz and $30$ degree half power beamwidth.\end{tabular}     \\ \bottomrule
\end{tabular}}
\end{table}

Throughout the measurement process, $300$ distinct in-phase and quadrature data, including 4G LTE and 5G NR signals for three transmission scenarios, are transmitted by the PT's SDR module. Then, the SU's SDR module records the spectrograms of the received signals, which are processed by the DL-based spectrum sensing models, namely Detectron2 and YOLOv7, for determining the signal types and extracting their spectral characteristics. The measurements are conducted in two distinct phases: \textit{(i)} the RIS is in off mode, specifying that the reflecting element phase shifts are set to $0^\circ$, and \textit{(ii)} the RIS operates in an optimized state, configuring the reflecting elements to maximize the received signal power at the SU. 

\begin{figure}[t]
   \centering
   \includegraphics[width=0.85\linewidth]{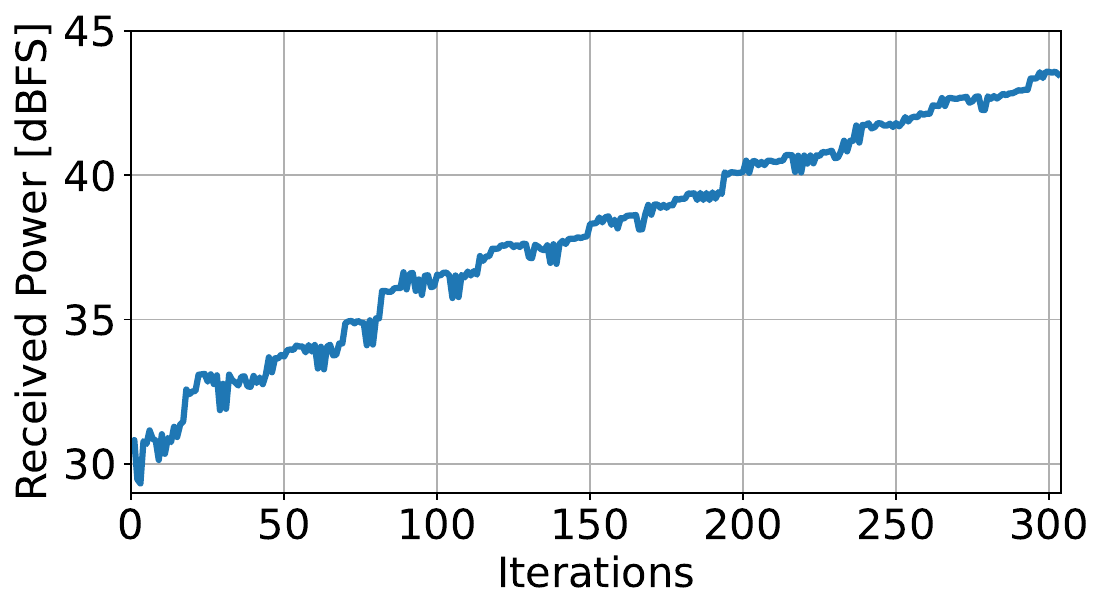}
   \caption{{\color{black}The received signal power during iterations.}}
   \label{fig:SSMeasurement} 
\end{figure}

\begin{figure}[t]
\centering
\subfloat[]{\label{fig:SpectrogramMeasurement:a} 
\includegraphics[width=0.48\linewidth]{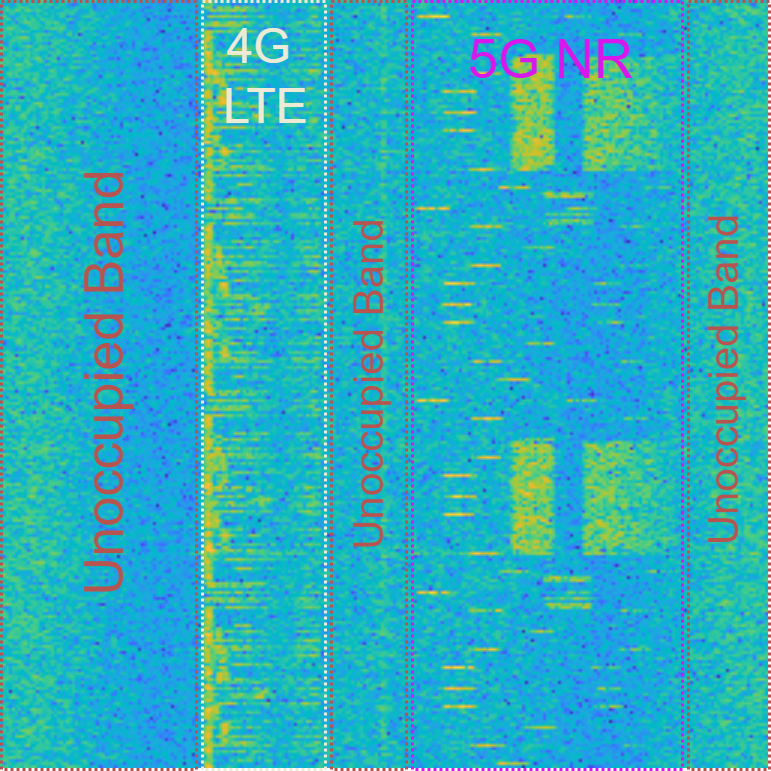}}
\subfloat[]{\label{fig:SpectrogramMeasurement:b} 
\includegraphics[width=0.48\linewidth]{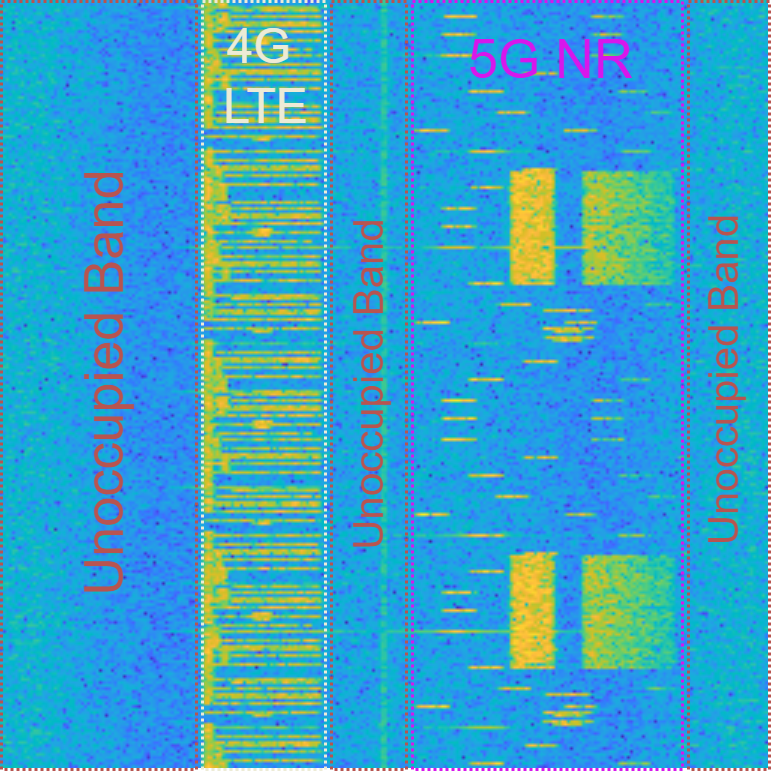}}
\caption{{\color{black}The spectrogram example of the received signal of both 4G LTE and 5G NR when the RIS is (a) off and (b) optimized.}}
\label{fig:SpectrogramMeasurement} 
\end{figure}

\begin{figure}[t]
\centering
\subfloat[]{\label{fig:SpectrogramExample:a} 
\includegraphics[width=0.48\linewidth]{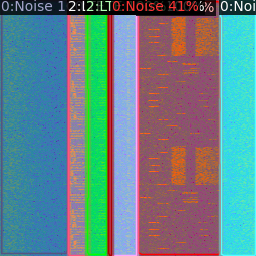}}
\subfloat[]{\label{fig:SpectrogramExample:b} 
\includegraphics[width=0.48\linewidth]{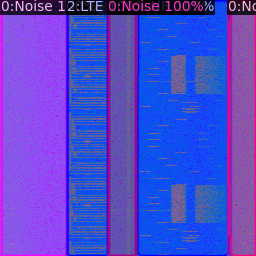}}\\[10pt]
\subfloat[]{\label{fig:SpectrogramExample:c} 
\includegraphics[width=0.48\linewidth]{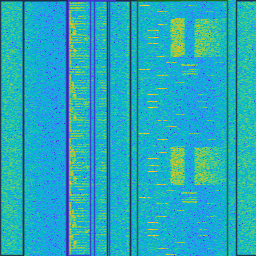}}
\subfloat[]{\label{fig:SpectrogramExample:d} 
\includegraphics[width=0.48\linewidth]{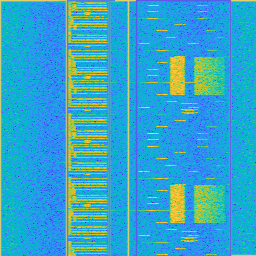}}
\caption{{\color{black}The output example of the detectors of (a) Detectron2, (c) YOLOv7 for the case of the RIS is off, and (b) Detectron2, (d) YOLOv7 for the case of the RIS is optimized.}}
\label{fig:SpectrogramExample} 
\end{figure}

The RIS is optimized through the iterative algorithm in \cite{kayraklik2022indoor} by updating the phase shift configurations of the reflecting elements one by one, where the received signal power of the SU is boosted by approximately $13$ dB after $300$ iterations. {\color{black} Fig. \ref{fig:SSMeasurement} illustrates the received signal power during the iterations of the iterative algorithm. Fig. \ref{fig:SpectrogramMeasurement} demonstrates the improvement in an example spectrogram of the received signal from the SU's SDR module during the measurements when the RIS is off and optimized by the iterative algorithm.} Therefore, the signals monitored by the SU become more apparent in the spectrogram representation.

{\color{black} The Detectron2's output of the spectrogram example given in Fig. \ref{fig:SpectrogramMeasurement} is illustrated in Fig. \ref{fig:SpectrogramExample}(a) and Fig. \ref{fig:SpectrogramExample}(b) for the RIS-off and RIS-optimized cases, respectively. Similarly, the YOLOv7's estimation of the spectrogram example is shown in Fig. \ref{fig:SpectrogramExample}(c) and Fig. \ref{fig:SpectrogramExample}(d) for the RIS-off and RIS-optimized scenarios, respectively. The results in Fig. \ref{fig:SpectrogramExample}(a) and Fig. \ref{fig:SpectrogramExample}(c) demonstrate that both the detectors fail to accurately estimate the boundaries of 4G LTE and 5G NR signals when the RIS is off since their SNRs are low and the entire signals are not visible in the spectrogram images. Also, both detectors cannot correctly detect all the unoccupied bands when the RIS is off. On the other hand, optimizing the RIS results in more apparent signals in the spectrogram images; hence, the estimations of the detectors become more accurate for both boundaries of 4G LTE and 5G NR and signal types.}

\begin{figure}[t]
\centering
\subfloat[]{\label{fig:SSMeaResults:a} 
\includegraphics[width=0.65\linewidth]{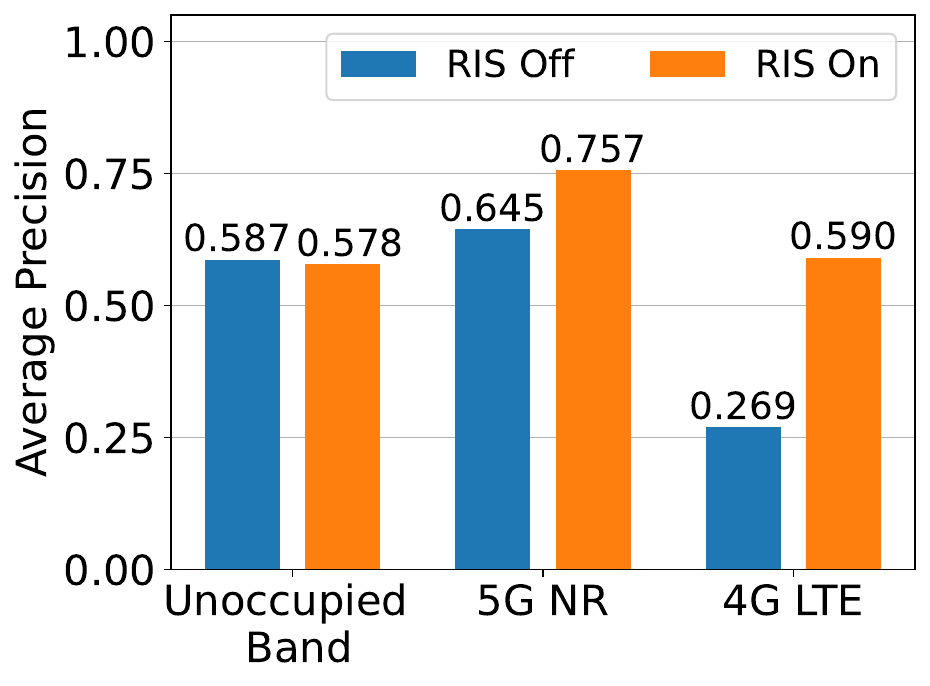}}\\[10pt]
\subfloat[]{\label{fig:SSMeaResults:b} 
\includegraphics[width=0.65\linewidth]{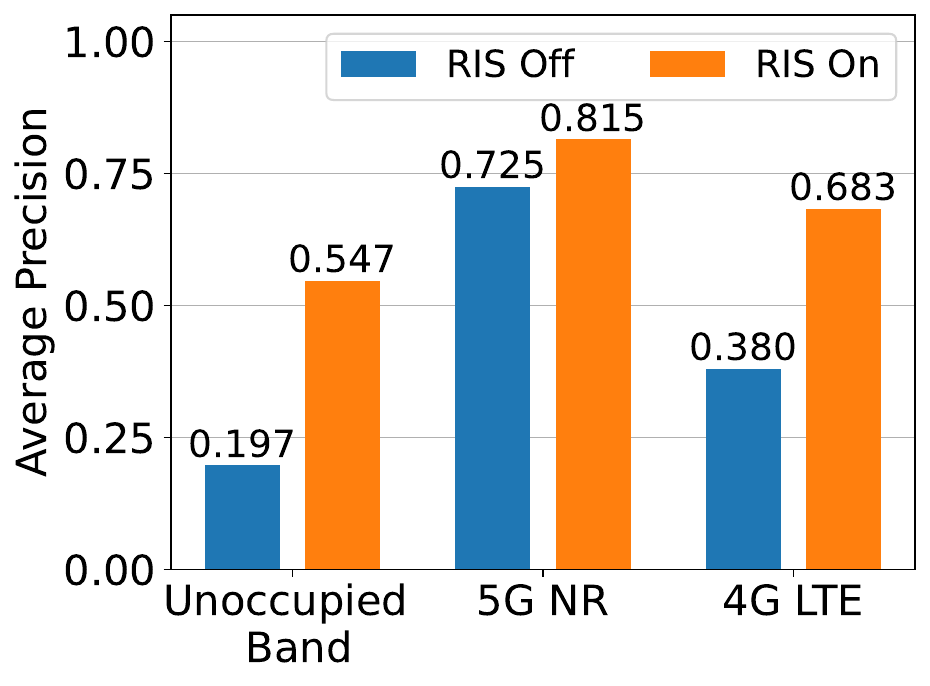}}
\caption{The AP results using the measurement dataset when the RIS is off and on for (a) Detectron2 and (b) YOLOv7.}
\label{fig:SSMeaResults} 
\end{figure}

Fig. \ref{fig:SSMeaResults} demonstrates the AP results of the detectors with the collected $300$ spectrograms from over-the-air by the SU's SDR for the two cases corresponding to the RIS being off and on. For the Detectron2, the optimized RIS configuration has the most significant impact on the 4G LTE signal by remarkably enhancing the AP from $0.269$ to $0.590$. Additionally, the AP of the 5G NR signal is increased from $0.645$ to $0.757$, while a slight reduction is observed for the unoccupied band. For the YOLOv7, the AP of the 4G LTE signal is raised from $0.380$ to $0.683$, the 5G NR signal's AP is boosted by $0.09$, and the highest improvement of $0.350$ is obtained for the unoccupied band portion of the signal. 
The performance of both detectors is improved for the signals of interest (i.e., 4G LTE and 5G NR) by introducing the RIS into the spectrum sensing system. Overall, YOLOv7 performs better than Detectron2 except for detecting the unoccupied band for both cases of RIS's operation modes. Therefore, utilizing YOLOv7 in the spectrum sensing process for determining the activities of the PT is more reasonable, as it demonstrates superior detection performance for the signals of interest. On the other hand, employing Detectron2 in the spectrum sensing system would be proper for the SU to determine the unused bands in the spectrum and utilize them.

\section{Conclusion}
An RIS-enhanced spectrum sensing system, where the secondary user monitors the activities of the PT with DL-based detectors, has been proposed. The state-of-the-art detectors, Detectron2 and YOLOv7, are utilized to estimate the signal's parameters, including signal type and location in the spectrum. These detectors are trained using the synthesized signal dataset of 4G LTE and 5G NR. The trained detectors are tested for their performance in two practical measurement environments in which the RIS is in off mode and optimized mode to focus the reflected beam toward the SU. The extensive measurement results demonstrate that the RIS significantly improves the average precision of the detectors in the spectrum sensing task.

\section{Acknowledgment}
We thank to StorAIge project that has received funding from the KDT Joint Undertaking (JU) under Grant Agreement No. 101007321. The JU receives support from the European Union’s Horizon 2020 research and innovation programme in France, Belgium, Czech Republic, Germany, Italy, Sweden, Switzerland, Türkiye, and National Authority TÜBİTAK with project ID 121N350. This work has been partially supported by the THULAB project. The work of Ertugrul Basar was supported by TUBITAK under Grant 121C254.

\bibliographystyle{IEEEtran}
\bibliography{main.bib}

\begin{thebibliography}{10}
\providecommand{\url}[1]{#1}
\csname url@samestyle\endcsname
\providecommand{\newblock}{\relax}
\providecommand{\bibinfo}[2]{#2}
\providecommand{\BIBentrySTDinterwordspacing}{\spaceskip=0pt\relax}
\providecommand{\BIBentryALTinterwordstretchfactor}{4}
\providecommand{\BIBentryALTinterwordspacing}{\spaceskip=\fontdimen2\font plus
\BIBentryALTinterwordstretchfactor\fontdimen3\font minus
  \fontdimen4\font\relax}
\providecommand{\BIBforeignlanguage}[2]{{%
\expandafter\ifx\csname l@#1\endcsname\relax
\typeout{** WARNING: IEEEtran.bst: No hyphenation pattern has been}%
\typeout{** loaded for the language `#1'. Using the pattern for}%
\typeout{** the default language instead.}%
\else
\language=\csname l@#1\endcsname
\fi
#2}}
\providecommand{\BIBdecl}{\relax}
\BIBdecl

\bibitem{Haykin_Sensing}
S.~Haykin, D.~J. Thomson, and J.~H. Reed, ``Spectrum sensing for cognitive
  radio,'' \emph{Proceedings of the IEEE}, vol.~97, no.~5, pp. 849--877, 2009.

\bibitem{Haykin_CR}
S.~Haykin, ``Cognitive radio: brain-empowered wireless communications,''
  \emph{IEEE Journal on Selected Areas in Communications}, vol.~23, no.~2, pp.
  201--220, 2005.

\bibitem{Basar_Access_2019}
E.~Basar, M.~Di~Renzo, J.~De~Rosny, M.~Debbah, M.-S. Alouini, and R.~Zhang,
  ``Wireless communications through reconfigurable intelligent surfaces,''
  \emph{IEEE Access}, pp. 116\,753--116\,773, Sep. 2019.

\bibitem{Wu_Tutorial}
Q.~Wu, S.~Zhang, B.~Zheng, C.~You, and R.~Zhang, ``Intelligent reflecting
  surface aided wireless communications: {A} tutorial,'' \emph{IEEE
  Transactions on Communications}, vol.~69, no.~5, 2021.

\bibitem{SimRIS_Mag}
E.~Basar and I.~Yildirim, ``Reconfigurable intelligent surfaces for future
  wireless networks: {A} channel modeling perspective,'' \emph{IEEE Wireless
  Communications}, vol.~28, no.~3, pp. 108--114, June 2021.

\bibitem{makarfi2021reconfigurable}
A.~U. Makarfi, R.~Kharel, K.~M. Rabie, O.~Kaiwartya, X.~Li, and D.-T. Do,
  ``Reconfigurable intelligent surfaces based cognitive radio networks,'' in
  \emph{2021 IEEE Wireless Communications and Networking Conference Workshops
  (WCNCW)}.\hskip 1em plus 0.5em minus 0.4em\relax IEEE, 2021, pp. 1--6.

\bibitem{wu2021irs}
W.~Wu, Z.~Wang, L.~Yuan, F.~Zhou, F.~Lang, B.~Wang, and Q.~Wu, ``{IRS}-enhanced
  energy detection for spectrum sensing in cognitive radio networks,''
  \emph{IEEE Wireless Communications Letters}, vol.~10, no.~10, pp. 2254--2258,
  2021.

\bibitem{lin2022intelligent}
S.~Lin, B.~Zheng, F.~Chen, and R.~Zhang, ``Intelligent reflecting surface-aided
  spectrum sensing for cognitive radio,'' \emph{IEEE Wireless Communications
  Letters}, vol.~11, no.~5, pp. 928--932, 2022.

\bibitem{nasser2022intelligent}
A.~Nasser, H.~A.~H. Hassan, A.~Mansour, K.-C. Yao, and L.~Nuaymi, ``Intelligent
  reflecting surfaces and spectrum sensing for cognitive radio networks,''
  \emph{IEEE Transactions on Cognitive Communications and Networking}, vol.~8,
  no.~3, pp. 1497--1511, 2022.

\bibitem{ge2022ris}
J.~Ge, Y.-C. Liang, S.~Li, and Z.~Bai, ``{RIS}-enhanced spectrum sensing: How
  many reflecting elements are required to achieve a detection probability
  close to 1?'' \emph{IEEE Transactions on Wireless Communications}, vol.~21,
  no.~10, pp. 8600--8615, 2022.

\bibitem{wu2023joint}
W.~Wu, Z.~Wang, Y.~Wu, F.~Zhou, B.~Wang, Q.~Wu, and D.~W.~K. Ng, ``Joint
  sensing and transmission optimization for {IRS}-assisted cognitive radio
  networks,'' \emph{IEEE Transactions on Wireless Communications}, 2023.

\bibitem{prasad2020downscaled}
K.~S.~V. Prasad, K.~B. D’souza, and V.~K. Bhargava, ``A downscaled
  faster-{RCNN} framework for signal detection and time-frequency localization
  in wideband {RF} systems,'' \emph{IEEE Transactions on Wireless
  Communications}, vol.~19, no.~7, pp. 4847--4862, 2020.

\bibitem{kayraklik2022application}
S.~Kayraklik, Y.~Alag{\"o}z, and A.~F. Co{\c{s}}kun, ``Application of object
  detection approaches on the wideband sensing problem,'' in \emph{2022 IEEE
  International Black Sea Conference on Communications and Networking
  (BlackSeaCom)}.\hskip 1em plus 0.5em minus 0.4em\relax IEEE, 2022, pp.
  341--346.

\bibitem{janu2022machine}
D.~Janu, K.~Singh, and S.~Kumar, ``Machine learning for cooperative spectrum
  sensing and sharing: A survey,'' \emph{Transactions on Emerging
  Telecommunications Technologies}, vol.~33, no.~1, p. e4352, 2022.

\bibitem{syed2023deep}
S.~N. Syed, P.~I. Lazaridis, F.~A. Khan, Q.~Z. Ahmed, M.~Hafeez, A.~Ivanov,
  V.~Poulkov, and Z.~D. Zaharis, ``Deep neural networks for spectrum sensing: A
  review,'' \emph{IEEE Access}, 2023.

\bibitem{wu2019detectron2}
Y.~Wu, A.~Kirillov, F.~Massa, W.-Y. Lo, and R.~Girshick, ``Detectron2,''
  \url{https://github.com/facebookresearch/detectron2}, 2019 (accessed July 20,
  2023).

\bibitem{wang2023yolov7}
C.-Y. Wang, A.~Bochkovskiy, and H.-Y.~M. Liao, ``Yolov7: Trainable
  bag-of-freebies sets new state-of-the-art for real-time object detectors,''
  in \emph{Proceedings of the IEEE/CVF Conference on Computer Vision and
  Pattern Recognition}, 2023, pp. 7464--7475.

\bibitem{boashash2015time}
B.~Boashash, \emph{Time-frequency signal analysis and processing: a
  comprehensive reference}.\hskip 1em plus 0.5em minus 0.4em\relax Academic
  press, 2015.

\bibitem{redmon2016yolo}
J.~Redmon, S.~Divvala, R.~Girshick, and A.~Farhadi, ``You only look once:
  Unified, real-time object detection,'' in \emph{Proceedings of the IEEE
  Conference on Computer Vision and Pattern Recognition}, 2016, pp. 779--788.

\bibitem{mathworks}
Mathworks, ``Spectrum sensing with deep learning to identify {5G} and {LTE}
  signals,''
  \url{https://www.mathworks.com/help/comm/ug/spectrum-sensing-with-deep-learning-to-identify-5g-and-lte-signals.html},
  2023.

\bibitem{greenerwave}
{Greenerwave}, ``Greenerwave 2022, 35 rue du {S}entier, 75002 {P}aris,
  {F}rance,'' \url{https://greenerwave.com/}, 2022 (accessed July 28, 2022).

\bibitem{kayraklik2022indoor}
S.~Kayrakl{\i}k, I.~Yildirim, Y.~Gevez, E.~Basar, and A.~G{\"o}r{\c{c}}in,
  ``Indoor coverage enhancement for {RIS}-assisted communication systems:
  Practical measurements and efficient grouping,'' in \emph{ICC 2023-IEEE
  International Conference on Communications}.\hskip 1em plus 0.5em minus
  0.4em\relax Rome, Italy: IEEE, 2023.

\end{thebibliography}
\end{document}